\documentclass{llncs}

\usepackage[all]{xy} 
\usepackage{amsmath,amssymb}

\usepackage{dsfont}
\title{Computing in Coq \\ with Infinite Algebraic Data Structures\thanks{Partially
 supported by Ministerio de Ciencia e Innovaci\'on, project
 MTM2009-13842-C02-01, and by European Commission FP7, STREP project ForMath.}}
\author{C\'esar Dom\'\i nguez \and Julio Rubio}
\institute{Departamento de Matem\'aticas y Computaci\'on, Universidad de La Rioja.\\
Edificio Vives, Luis de Ulloa s/n, E-26004 Logro\~ no (La Rioja,
 Spain). \email{cesar.dominguez,julio.rubio@unirioja.es}\thanks{The final publication of this paper is
   available at www.springerlink.com}}

\begin{document}
\maketitle
\begin{abstract}
Computational content encoded into constructive type theory proofs can be
used to make computing experiments over concrete data structures.
In this paper, we explore this possibility when working in Coq with
chain complexes of infinite type (that is to say, generated by infinite sets)
as a part of the formalization of a hierarchy of homological algebra structures.
\end{abstract}

\begin{keywordname}
Theorem proving, formal methods, computer algebra, program verification.
\end{keywordname}
\sloppy
\section{Introduction}
One main feature of constructive type theory, through the well-known
Curry-Howard isomorphism, is the equivalence between proving and programming.
This is clearly one of the advantages of Coq~\cite{BC04} with respect to other proof
assistants, like Isabelle/HOL~\cite{NIPAWE02}.
This characteristic is the base of reflective tactics, pioneered by S. Boutin~\cite{Boutin97},
and successfully used, for instance, in~\cite{Gonthier,Mahboubi07}.

Computing can play another role when formalizing a proof. It can
be useful, for example, to check some conjecture over concrete
cases. When dealing with standard data structures (as lists, trees,
and the like), these experiments can be done in a parallel line by
programming the tests in Java, C, or any other programming language.
If infinite data structures occur, programming them is a more
delicate task, and it can be rewarding to keep a tighter link among
programs and specifications.

Infinite data structures, presented as coinductive objets as streams, have been
dealt with in the theorem proving literature (see~\cite{BC04} for instance). In this work, we
undertake another via to manage the infinity, working with algebraic structures of
infinite type (that is to say, generated by infinite sets)~\cite{BSM}.
We report in this paper on an experiment of this nature, in the area
of homological algebra. It is well-known that homological
information is not computable over general (infinite type) chain
complexes (see~\cite{IJCM}). For instance, if $(C,d)$ is an
\emph{acyclic} chain complex, and $x \in C_n$ is a \emph{cycle}
(this means $d_n(x)=0$), then \emph{there exists} $z \in C_{n+1}$
such that $d_{n+1}(z)=x$ (that is, $x$ is a \emph{boundary}). But if
$C_{n+1}$ is a free module of infinite type, and no other
information is available, there is no general algorithm computing a
pre-image $z$ of $x$.

Sergeraert's effective homology~\cite{Advances} is a theory allowing solving large classes
of problems of this sort, even in the infinite dimensional case. This paper continues our
previous work in translating Sergeraert's ideas to theorem provers~\cite{JAR,FAC,Eurocast2009,ALR07}, with the
aim of formalizing this part of algorithmic mathematics and, more importantly, of applying
formal methods to the study of the Kenzo system~\cite{Kenzo}
(a Common Lisp program developed by Sergeraert to implement effective homology algorithms). The first important milestone in this area was the mechanized proof in the Isabelle/HOL proof assistant
of the Basic Perturbation Lemma (BPL), published in~\cite{JAR}. This formal proof was carried out in the Higher
Order Logic (HOL) built on top of Isabelle, and therefore extracting programs from it was not a simple task. The
findings on this topic were reported in~\cite{FAC}. A different approach
is being carried out by T. Coquand and A. Spiwack \cite{CS07} who are using Coq to model a part of
Category Theory, and then trying to obtain a BPL proof in this larger context.

The data structures of effective homology are organized in two
layers (as algebraically modeled in~\cite{AAECC,Rairo}): the first
layer is composed of algebraic data structures (chain complexes,
simplicial sets, \ldots) and the second one of standard data
structures (lists, trees, \ldots) which are representing
\emph{elements} of data from the first layer. Infinite type data
structures appear only in the first layer. Computing in this first
layer can be done in an abstract way, and it is equivalent in Coq to
proving theorems. For example, a theorem stating ``the direct sum of
two chain complexes is a chain complex'' contains an algorithm
constructing the mentioned direct sum. Coq can deal with this
structure, no matter whether it is of finite or infinite type. But
actual computations really take place \emph{within} algebraic
structures of the first layer. To compute with Coq in this sense has
no advantage of being any more direct. It is needed to construct
concrete instances of chain complexes and other possibly infinite
algebraic data structures. Then we must build concrete elements
(second layer) of these particular structures, and finally put to
work the algorithms abstractly described in the first layer.

In this paper we discuss this procedure in a case related to the
effective homology of the cone of a chain complex morphism. This
formalization was part of the implementation in Coq of the algorithm
computing the effective homology of a bicomplex
(see~\cite{preprint}). Now, we use the computing capabilities of Coq
to explore whether some concrete cones are acyclic or not, as a
previous step to proving a general property.

The paper is organized as follows. Section 2 contains some
preliminaries on algebraic structures, both in Mathematics and in
Coq. Section 3 describes the formalization in Coq of the algorithm computing the
effective homology of a cone, in a way that slightly generalizes our
previous work in~\cite{preprint}. Concrete Coq instances of chain
complexes of infinite type are introduced in Section 4. Then
explicit calculations with elements are presented in Section 5,
using Coq as a computing tool to check some conjectures. The paper
ends with conclusions, future work, and the bibliography.
The Coq source files are available at \\ \texttt{https://esus.unirioja.es/psycotrip/archivos\_documentos/CCIADS.zip}.

\section{Algebraic Data Structures in Coq}\label{S:Preliminaries}
In this section we introduce the algebraic structures which support
our constructions. They include chain complexes, chain
complex morphisms, and reductions and effective homologies of
chain complexes. The formalization in Coq of these structures are also
described.

We assume as known the notions of \emph{ring}, \emph{module} over a
ring and \emph{module morphism} (see~\cite{Jacobson} for instance).
A ring $R$ commutative and with unity is fixed all through the
paper, and modules are supposed to be \emph{left} $R$-modules.

We have built these basic structures in Coq using records called
\texttt{Ring}, \texttt{Module} and \texttt{ModHom}, respectively.
 They are based on the ones included in CoRN~\cite{Corn}
(but simplifying them: basically eliminating the apartness relation included in
setoids which is not used by us, since we are working in a \emph{discrete}
mathematics setting). Besides, further constructions as for instance the addition or the
composition of module morphisms are defined, and are represented
using the infix notation \texttt{[+h]} or \texttt{[oh]},
respectively.

A \emph{free $R$-module generated over a set $B$} is the module
$R[B]$ whose elements are linear combinations with elements of $B$
as generators. The addition and the external product by elements of
$R$ are defined in the natural way. Since we are planning to work in
a constructive logic setting, it is convenient to define a
\emph{free} module as one module $M$ where an explicit isomorphism
is known between $M$ and $R[B]$ (the set of generators $B$ must also
be explicitly given). If $B$ is finite, the free module is said
\emph{of finite type}.

The formalization of free modules in Coq follows the ideas given by
L. Pottier in the Coq contributions web page~\cite{Coq}. There, a
definition can be found of a module built by freely generation
from a basis, which is given by a \emph{setoid} (\emph{i.e.} a set with an equality, usually denoted by \texttt{[=]}),
 using the module
operations. If we call $B$ the basis setoid, this is representing
the mathematical structure $R[B]$ introduced above.
Then, our formalization of free modules
 consists of a record with a module and an
explicit isomorphism to such a freely generated module. In order to deal with finite
sets in a constructive type theory, more care is needed.
For instance, several alternatives for defining  finite sets
in a constructive logic are included in~\cite{CS09}.
Finite algebraic structures have also been
implemented in Coq in~\cite{GMRTT07} as a first milestone of a
long-term effort to formalize the Feit-Thompson theorem.  Our formalization is the following.
Given a natural number $k \in \mathds{N}$,
let us denote $FS(k)$ the (finite) setoid $\{0,1,\ldots,k-1\}$ (with the Leibniz equality).
We consider a setoid $B$ as \emph{finite} if it is endowed with a natural number $k \in \mathds{N}$ and an explicit
bijection to $FS(k)$. Then, a free module of finite
type is a free module, but we impose that the generator set
is \emph{equal} (in the Coq internal sense) to $FS(k)$.

We concentrate ourselves in the sequel on \emph{free} modules, since
it is the unique kind of modules dealt with in the Kenzo system
\cite{Kenzo}.

We are ready to introduce the first \emph{graded} concept, needed in
Homological Algebra and Algebraic Topology.

\begin{definition}\label{D:graded_module}
A \emph{graded module} $M$ is a family of $R$-modules indexed by the integer numbers
$(M_{i})_{i \in \mathds{Z}}$. A graded module is \emph{free} (or
\emph{free of finite type}) if $M_{i}$ is free (free of finite type, respectively) for all $i \in
\mathds{Z}$. If $x\in M_{i}$, the index $i$ is called \emph{degree} of the element $x$.
\end{definition}

\begin{definition}
\label{D:differential} Given a graded module $M$ a
\emph{differential operator} $d$ on $M$ is a family of module
morphisms $(d_{i}\colon M_{i+1} \to M_i)_{i \in \mathds{Z}}$ such
that $d_i \circ d_{i+1} = 0$ for all $i \in
\mathds{Z}$.
\end{definition}


\begin{definition}
\label{D:chain_complex} A \emph{chain complex} is a pair $CC=(M, d)$
where $M$ is a graded module and $d$ a differential operator on $M$.
A chain complex is called \emph{free} (or \emph{free of finite
type}) when its underlying graded module is free (free of finite
type, respectively).
\end{definition}

Chain complexes have a corresponding notion of morphism.
\begin{definition}
A \emph{chain complex morphism} (or, simply, a \emph{chain morphism}) $f\colon CC \to CC'$ between two chain complexes $CC=(M,
d)$ and $CC'=(M', d')$ is a family of module
morphisms $(f_i\colon M_i \to M'_i)_{i \in \mathds{Z}}$ such that
$f_i \circ d_i=d'_i \circ f_{i+1}$ for all $i \in \mathds{Z}$.
\end{definition}

Given a ring \texttt{R: Ring},  a graded module can be formalized in
Coq with the following \emph{dependent} type: \texttt{Z -> Module R}, which accurately represents
a family of modules indexed by the integer numbers. Then,
a (free) chain complex can be formalized in Coq using the
following record structure:
 {\small\begin{verbatim}
 Record ChainComplex: Type:=
  {GrdMod:> Z -> FreeModule R;
   Diff: forall i:Z, ModHom (R:=R) (GrdMod (i + 1)) (GrdMod i);
   NilpotencyDiff: forall i:Z, (Nilpotency (Diff i)(Diff (i + 1))}. \end{verbatim}}
\noindent where the nilpotency property is defined by
{\texttt{Nilpotency(g:ModHom B C) (f:ModHom A B):= forall a: A,
((g[oh]f)a)[=]Zero}.}

In a similar way, given two chain complexes \texttt{CC1 CC2:
ChainComplex R}, a chain complex morphism \texttt{ChainComplexHom}
is represented as a record with a family of module morphisms
\texttt{GrdModHom:>forall i:Z,ModHom(CC1 i)(CC2 i)} which
commutes with the chain complex differentials.

Now, the central definition in effective homology theory: \emph{reduction}.
A reduction establishes a link between a ``big'' chain complex, called
\emph{top complex}, and a smaller one, called \emph{bottom complex}, in such a way
that if all the homological problems are solved in the bottom complex, then it is
the same in the top one.

\begin{definition}\label{D:reduction}
A \emph{reduction} is a 5-tuple $(TCC, BCC, f, g, h)$
$$
  \xymatrix@C=4pc{TCC\ar@/^/[r]^{f}\ar@/4pc/@(ul,dl)_{h}& \ar@/^/[l]^{g}BCC
  }
$$ where
$TCC=(M, d)$ and $BCC=(M', d')$ are chain complexes (named
\emph{top} and \emph{bottom} chain complex), $f\colon TCC \to BCC$
and $g\colon BCC \to TCC$ are chain morphisms, $h=(h_i\colon
M_i \to M_{i+1})_{i \in \mathds{Z}}$ is a family of module morphisms
(called \emph{homotopy operator}), which satisfy the following
properties for all $i \in \mathds{Z}$: \vskip 4pt
\begin{enumerate}
\item  $f_i \circ g_i = id_{M'_{i}}$
\item $d_{i + 1} \circ h_{i+1} + h_i \circ d_i + g_{i+1} \circ f_{i+1} = id_{M_{i+1}}$
\item $f_{i+1} \circ h_{i}= 0 $
\item $h_i \circ g_i = 0 $
\item $h_{i + 1} \circ h_i = 0 $
\end{enumerate}
\end{definition}

And now, the relevant case. In a free chain complex  \emph{of finite type}
the homological problems can be solved algorithmically in a simple way
(at least in cases where the ring $R$ allows one to diagonalize matrices
over $R$; this includes the case $R = \mathds{Z}$, the most important one
in Algebraic Topology; see~\cite{BSM}). Thus, if from a chain complex
(possibly of infinite type) we can get a reduction to a chain complex of
finite type, the homological problem is solved for the initial complex.
This is the strategy followed in the Kenzo system. And it is the very notion
of chain complex with \emph{effective homology}.

\begin{definition}
A chain complex $CC$ is with \emph{effective homology} if it is free
and it is endowed with a reduction where $CC$ itself is the top chain complex and the bottom chain complex is free of finite type.
\end{definition}


Given a chain complex \texttt{CC1:
ChainComplex R}, a homotopy operator
is represented in Coq as a family of module morphisms
\texttt{HomotopyOperator:= forall i: Z, ModHom(C1 i)(C1(i + 1))}.
The reduction notion is then formalized as a record
\texttt{Reduction} with two chain complexes \texttt{topCC:ChainComplex R},
\texttt{bottomCC:ChainComplex R} and three
morphisms \texttt{f\_t\_b:ChainComplexHom topCC bottomCC},
\texttt{g\_b\_t:ChainComplexHom bottomCC topCC}, \texttt{h\_t\_t:HomotopyOperator topCC}.
Besides, five fields  representing the five
reduction properties are included. For instance, the field which
corresponds to the second property is:
\texttt{rp2: homotopy\_operator\_property f\_t\_b g\_b\_t h\_t\_t} with:
 {\small\begin{verbatim}
 Definition homotopy_operator_property:= forall(i: Z)(a: C1(i+1)),
  (((Diff C1(i+1))[oh]h(i+1))[+h](h i[oh](Diff C1 i))[+h]
       (g(i+1)[oh]f(i+1))) a [=] a. \end{verbatim}}
Some comments on these Coq definitions are needed. Why are the
elements in this definition considered to be on the \texttt{i+1}-th
degree and not on the \texttt{i}-th degree, as it is the usual
definition of reduction? The same decision was previously taken when
the definition of differential was introduced. It is clear that as
we are considering the definition for all the integers, both
definitions are equivalent. But, a Coq technical problem is easily
avoided thanks to our definition. We are going to focus our
attention on the \mbox{\texttt{(h i[oh](Diff C1 i))}} component of
the definition. The differential takes an element in degree
\texttt{i+1} and obtains an element in degree \texttt{i} which is
translated to a component in degree \texttt{i+1} by the homotopy
operator. If we consider the \emph{mathematically equivalent}
definition, considering the differential defined from degree
\texttt{i} to \texttt{i-1}, then the corresponding component would
be  \texttt{(h(i-1)[oh](Diff C1 i))}. In this composition, the
differential takes an element in degree \texttt{i} and returns an
element in degree \texttt{i-1}, which is now translated to a
component in degree \texttt{i-1+1}. In Coq this element is
\emph{equal} but is not \emph{convertible} to \texttt{i}. So, we
will obtain a Coq type error from this sum of morphisms. A
\emph{transition} function between equal but not directly
convertible types (which it is essentially an identity between
types) can be introduced allowing us to overcome this
drawback\footnote{We acknowledge T. Coquand for the suggestion of
this idea.}.


The concept of  \emph{free of finite type chain complex} is then
obtained in Coq as a specialization of the \emph{chain complex} structure:
simply adding that the family of modules are free modules of finite
type. In a similar way it is formalized the concept of \emph{effective
homology} as a specialization of the \emph{reduction} structure by
declaring the \emph{bottomCC} is of finite type.

\section{Effective Homology of the Cone in Coq}
In this section we first define the notion of the cone of a chain
complex morphism. Then, the main result that we are going to deal
with is stated: the effective homology of a cone. We also show how
this theorem can be proved in Coq.

\begin{definition}\label{D:cone} Given a pair of chain complexes $CC=((M_{i})_{i \in
\mathds{Z}}, (d_{i})_{i \in \mathds{Z}})$ and $CC'=((M'_{i})_{i \in
\mathds{Z}}, (d'_{i})_{i \in \mathds{Z}})$ and a chain complex
morphism  $\alpha\colon CC \to CC'$, the \emph{cone} of $\alpha$,
denoted by $Cone(\alpha)$,  is a chain complex $((M''_{i})_{i \in
\mathds{Z}}, (d''_{i})_{i \in \mathds{Z}})$ such that, for each $i
\in \mathds{Z}$, $M''_i= M_i\oplus M'_{i+1}$ and $d''_{i}(x,x')=
(-d_i(x),d'_{i+1}(x') + \alpha_{i+1}(x))$ for any $x\in M_{i+1}$ and
$x'\in M'_{i+2}$.
\end{definition}

Now, the theorem which determines the effective homology of a cone
can be stated.

\begin{theorem}\label{T:conereduction}
Given two reductions $r=(TCC, BCC, f, g, h)$ and $r'=(TCC', BCC',
f', g', h')$ and a chain morphism $\alpha\colon TCC \to TCC'$
between their top chain complexes, it is possible to define a
reduction $r''=(Cone(\alpha), BCC'', f'', g'', h'')$ with
$Cone(\alpha)$ as top chain complex and:
\begin{itemize}
\item $BCC''=Cone(\alpha')$ with $\alpha'\colon BCC \to BCC'$ defined by $\alpha'=f'\circ\alpha\circ g$
\item $f''= (f,f'\circ\alpha\circ h + f')$,  $g''= (g,-h'\circ\alpha\circ g + g')$, $h''= (-h,h'\circ\alpha\circ h + h')$
\end{itemize}
$$
  \xymatrix@C=4pc@R=2pc{TCC\ar@/^/[rr]^f\ar@/4pc/@(r,u)_h \ar[d]_{\alpha}& & BCC \ar@/^/[ll]^g \ar@{-->}[d]_{\alpha'}\\
  TCC'\ar@/^/[rr]^{f'}\ar@/4pc/@(r,d)^{h'} & & BCC' \ar@/^/[ll]^{g'}\\
    }
$$
Besides, if $TCC$ and $TCC'$ are objects with effective homology
through the reductions $r$ and $r'$, then  $Cone(\alpha)$ is an
object with effective homology through $r''$.
\end{theorem}

In~\cite{preprint} we formalized in Coq the effective homology of a
bicomplex. That result can be considered as a generalization of the
previous theorem to an infinite (indexed by the natural numbers)
family of reductions. Nevertheless, in order to obtain it, the chain
complexes must be positive, \emph{i.e.}, with null components in the
negative indexes (or, in other equivalent presentation, indexed by the
natural numbers). In this paper, we have not this constraint since
we work with a general definition of chain complex, with modules indexed
by integer numbers.

The formalization of Theorem~\ref{T:conereduction} in Coq is obtained as follows.
Given two chain complexes \texttt{CC0 CC1: ChainComplex R} and a
chain complex morphism \texttt{f: ChainComplexHom CC1 CC0}, the cone
of this morphism is a chain complex with family of modules
\texttt{ConeGrdMod:= fun i: Z => Sum\_FreeModule (CC1 i) (CC0(i+1))}
(with the direct sum of free modules \texttt{Sum\_FreeModule}
defined in a natural way) and with differential operator defined as
follows:
 {\small\begin{verbatim}
 Definition ConeDiffGround:= fun (i: Z)(ab:(ConeGround (i+1))) =>
  ([--](Diff CC1 i(fst ab)), ((Diff CC0(i+1))(snd ab)[+]f(i+1)(fst ab))).\end{verbatim}}
\noindent It is not difficult to prove that these functions define a
module morphism which satisfies the differential condition. This
last property allows one to build the cone chain complex associated
to a chain complex morphism: \texttt{Cone(f)}.

Given now two reductions \texttt{r1 r2: Reduction R} and a chain
complex morphism between their top chain complexes \texttt{alpha:
ChainComplexHom(topCC r1) (topCC r2)}, it is possible to define a
chain complex morphism \texttt{alpha'} between the bottom chain
complexes through the function \texttt{alpha'':= fun n:
Z => (f\_t\_b r2 i)[oh](alpha i)[oh](g\_b\_t r1 i).}

The first part of Theorem~\ref{T:conereduction} is proved if we build a
reduction between \texttt{Cone(alpha)} and \texttt{Cone(alpha')}.
The first chain complex morphism of the reduction is defined in the
following way:
{\small\begin{verbatim}
 Definition f_cone_reductionGround:
  forall i: Z, (Cone alpha) i -> (Cone alpha') i:=
   fun (i: Z)(ab: (Cone alpha) i) => ((f_t_b r1 i) (fst ab),
    (((f_t_b r2 (i+1)) [oh] (alpha (i+1)) [oh] (h_t_t r1 i)) (fst ab)) [+]
      (f_t_b r2 (i+1)) (snd ab)). \end{verbatim}}
Analogous definitions are provided for the two other morphisms
of the reduction. Then we state Coq lemmas for the reduction
properties on these morphisms. The proof of these lemmas consists in
applying mainly equational reasoning over setoid equalities,
following closely the \emph{paper and pencil} proof.  It allows
building the reduction of a cone: \texttt{ConeReduction(alpha)}.

Finally, given two effective homologies \texttt{r1 r2:
EffectiveHomology R} and a chain complex morphism \texttt{alpha} between their top
chain complexes, \texttt{ConeReduction(alpha)}
is directly a reduction of the cone. Then, in order to define
an effective homology for the cone it remains to prove that the
bottom free chain complex of this reduction is free of finite
type. It is easily obtained in Coq since the direct sum of free chain
complex of finite type is free of finite type.

\section{Instances of Chain Complexes of Infinite Type} \label{S:Instances}
A working representation in a proof assistant of the concepts included in previous sections
has to be sound, but also needs to be useful. The second feature can
be shown by formally proving some results. This was the purpose of the previous section.
The first feature can be illustrated by providing instances of the representations,
that accurately reflect usual mathematical entities.  This is the
aim of this section which includes different instances of all
the previous structures.

First, we  define some elementary instances which will act as building blocks
for more elaborated constructions. The first example is the null
free module $M^{(0)}$ (\emph{i.e.}, a module with the unit as
unique element). This is indeed a free module of finite type,
generated by the setoid with zero elements. Then, a null free  chain
complex can be defined $CC^{(0)}=((M^{(0)})_{i \in \mathds{Z}},(d^{(0)})_{i \in \mathds{Z}})$
(\emph{i.e.}, with the previous module in each degree and the null
differential). This chain complex can be also built as a free chain complex of finite type
$FCC^{(0)}$, defined from the corresponding free  module of finite type.
Obviously, a trivial effective homology for this chain complex can be
defined.

Another basic example is the free module of the integers
$\mathds{Z}$ (over the ring of integers) which we denote in Coq by
\texttt{ZFreeModule}. This module can be also implemented
as a module of finite type,
\texttt{ZFinFreeModule}, generated by the setoid with only one
element. Then, an example of free chain complex is
\mbox{$CC^{(1)}=((M^{(1)})_{i \in
\mathds{Z}},(d^{(1)})_{i \in \mathds{Z}})$} with
$(M^{(1)})_i = \mathds{Z}$, $\forall i \in
\mathds{Z}$, and $(d^{(1)})_{i}\colon \mathds{Z}\to
\mathds{Z}$ such as $(d^{(1)})_i(x)= 2 * x$ if $i$ is
even and $(d^{(1)})_{i}(x)= 0$ otherwise:
$$\xymatrix@C=2pc@R=0.25pc{\dots& \mathds{Z}\ar[l]_{0} & \mathds{Z}\ar[l]_{\times 2} & \mathds{Z}\ar[l]_{0} & \mathds{Z}\ar[l]_{\times 2} &\mathds{Z}\ar[l]_{0} & \ar[l]_{\times 2}\dots\\
\txt{\small{degree}}& \txt{\small{-2}} & \txt{\small{-1}} & \txt{\small{0}} & \txt{\small{1}} &\txt{\small{2}} & \\
}
$$
\noindent The Coq formalization of the required differential is
obtained through the functional type \texttt{fun i: Z => if
(Zeven\_bool i) then x2\_ModHom else (ModHom\_zero ZFreeModule
ZFreeModule)}. It is easy to prove that this morphism satisfies the
nilpotency condition. A similar free  chain complex of finite type
$FCC^{(1)}$ can be defined using the corresponding
family of free modules of finite type. Besides, we can define a
trivial effective homology between both complexes that we name
\texttt{Id\_Z\_2x\_0\_EffectiveHomology}:
$$
  \xymatrix@C=4pc{CC^{(1)}\ar@/^/[rr]^{id}\ar@/4pc/@(r,u)_{0} & & FCC^{(1)} \ar@/^/[ll]^{id}
  }
$$

The previous examples are  chain complexes of \emph{finite type}, since
the modules are free of finite type (in that case with zero or one
generator). An example of a free module of \emph{infinite type} is
$\mathds{Z}[\mathds{N}]$, the free module generated by the natural numbers
(over the ring of integer numbers)  which we denote
in Coq by \texttt{Z\_nat\_FreeModule}. It is defined by taking as free module the
one freely
generated from the setoid denoted in Coq by \texttt{nat\_as\_Setoid} (that is to say,
the setoid of natural numbers with the Leibniz equality). The
definition is then completed with the same module as module representation and the identity as isomorphism
between them. To keep notations clear, the generator $i$ of $\mathds{Z}[\mathds{N}]$ will be denoted
by $x_i$, $\forall i \in \mathds{N}$.

Now, a  chain complex of infinite type
$CC^{(2)}=((M^{(2)})_{i \in
\mathds{Z}},(d^{(2)})_{i \in \mathds{Z}})$ is
built where $(M^{(2)})_i =
{\mathds{Z}[\mathds{N}]}$, $\forall i \in \mathds{Z}$, and
$(d^{(2)})_i\colon \mathds{Z}[\mathds{N}]\to
\mathds{Z}[\mathds{N}]$ defined on generators (and then extended to
all elements by freely generation) in the following way: if $i$ is
even, $(d^{(2)})_i(x_j)= x_j$ if $j$ is even and
$(d^{(2)})_i(x_j)= 0$ otherwise; and if $i$ is
odd, $(d^{(2)})_i(x_j)= 0$ if $j$ is even and
$(d^{(2)})_i(x_j)= x_j$ otherwise. This
differential on generators can be illustrated with the following
diagram:

$$\xymatrix@C=2pc@R=0.7pc{ \mathds{Z}[\mathds{N}]& \mathds{Z}[\mathds{N}]\ar[l]^{(d^{(2)})_i}_{\txt{\scriptsize{i even}}}&  \mathds{Z}[\mathds{N}]& \mathds{Z}[\mathds{N}]\ar[l]^{(d^{(2)})_i}_{\txt{\scriptsize{i odd}}}\\
 x_0 &  x_0\ar@{|->}[l] & 0 &  x_0\ar@{|->}[l]\\
 0 & x_1 \ar@{|->}[l]& x_1 & x_1 \ar@{|->}[l]\\
 x_2 & x_2 \ar@{|->}[l]& 0 & x_2 \ar@{|->}[l]\\
 0 & x_3 \ar@{|->}[l]&  x_3 & x_3 \ar@{|->}[l]\\
\dots & \dots & \dots & \dots \\
}
$$
\noindent This chain complex is named in our representation
\texttt{Z\_nat\_ChainComplex}. Its differential can be easily defined using
auxiliary functions as \texttt{fun n: nat\_as\_Setoid =>
if even\_bool n then Var \_ n else Unit \_ \_}. Here, we are using the \texttt{Unit}
notation for the null element as in L. Pottier' s development. It is not difficult to prove that this morphism
satisfies the nilpotency condition (in other words, it is really a differential).

Now, it is possible to define a homotopy operator
$h^{(2)}$ on $CC^{(2)}$ built
on generators in the same way as the previous differential (but,
defined from an element in the module at degree $i$ to an element in
the module at degree $i+1$). Obvious morphisms allow us to complete
an effective homology from this last free chain complex to the
null free chain complex of finite type $FCC^{(0)}$. This last effective homology
proves that $CC^{(2)}$ is acyclic.

In order to define a more interesting effective homology we define
the free chain complex $CC^{(1)}\oplus CC^{(2)}$ obtained from the direct
sum of the two
previous chain complexes. Then, it is easy to define an effective
homology \texttt{Z\_x\_Z\_nat\_EffectiveHomology}:
$$
  \xymatrix@C=4pc{CC^{(1)}\oplus CC^{(2)}\ar@/^/[rr]^{\pi_1}\ar@/4pc/@(r,u)_{(0,h^{(2)})}& & \ar@/^/[ll]^{(id, 0)}FCC^{(1)}
  }
$$
\noindent where $\pi_1$ is the canonical projection in the first component.

Finally, we consider a free chain morphism between the top chain
complexes of  \texttt{Z\_x\_Z\_nat\_EffectiveHomology} and
\texttt{Id\_Z\_2x\_0\_EffectiveHomology} again through the canonical projection in
the first component:
$$
  \xymatrix@C=4pc@R=3pc{CC^{(1)}\oplus CC^{(2)}\ar@/^/[rr]^{\pi_1}\ar@/4pc/@(r,u)_{(0,h^{(2)})} \ar[d]_{\pi_1}& & \ar@/^/[ll]^{(id, 0)}FCC^{(1)}\ar[d]_{\alpha'}\\
  CC^{(1)}\ar@/^/[rr]^{id}\ar@/4pc/@(r,d)^{0} & & FCC^{(1)} \ar@/^/[ll]^{id}
    }
$$

Then, we can obtain in Coq the cone of this morphism and the effective
homology associated to it, named
\texttt{Example\_Cone\_EffectiveHomology}, as a particular instance
of our general result developed in the previous section:
$$
  \xymatrix@C=4pc{Cone(\pi_1)\ar@/^/[rr]^{f^{Ex}}\ar@/4pc/@(r,u)_{h^{Ex}}& &Cone(\alpha') \ar@/^/[ll]^{g^{Ex}}
  }
$$
\noindent We will use
this effective homology instance to make concrete computations in Coq in the following
section.

\section{Computing with Infinite Data Structures in Coq}
Working in the Coq constructive type theoretic setting  allows us to obtain from proofs directly computable terms.
In the previous section we obtained instances of meaningful examples of all
our data structures, so we can now make calculations with them through the associated
algorithms (which have been proved correct in Coq).
In particular we can make computations within instances of chain complexes
of infinite type.

We will use the \texttt{vm\_compute} Coq tactic for evaluating
terms. It computes the goal using the optimized call-by-value
evaluation bytecode-based virtual machine~\cite{Coq}. Another option
consists in using the Coq extracting code mechanism. Nowadays, the
functional languages available as output in Coq are OCaml, Haskell
and Scheme~\cite{Letouzey08}. This extracted code should be, in
principle, efficient but the presence of dependent types makes it
complicated, at least in the Haskell case. Being Scheme a kind of
Lisp, its dynamical typing style should be more convenient from this
point of view in order to be our target language in which extracts
our code.  Nevertheless it seems to be the least developed frame
(see~\cite{Coq} again). Since Kenzo is implemented in Common Lisp it
is clear that the problems encountered with Scheme are important for
us if we want to extract code which was directly comparable with the Kenzo code.
We do not follow this line in this paper. We explore rather the
possibilities of the \emph{internal} execution of Coq terms.

We are going to choose as an example the top chain complex of
\texttt{Example\_Cone\_EffectiveHomology}, \emph{i.e.}
$Cone(\pi_1)$. This is an example of chain complex of \emph{infinite
type}. For instance, we want to compute its differential applied to
the element $(5, 7*x_4 + 8*x_0, 3)$ at degree $2$. Since the module
at degree $2$ of the cone (and, in fact, at \emph{any} degree) is
$\mathds{Z} \oplus \mathds{Z}[\mathds{N}] \oplus \mathds{Z}$, the
element $(5, 7*x_4 + 8*x_0, 3)$ has a component in each module. The
first and third components appear simply as integers, because
$\mathds{Z}$ is considered a free module over a singleton which is
skipped. On the contrary, elements in the second component are true
combinations in $\mathds{Z}[\mathds{N}]$ with generator $x_i$
(recall our convention of naming $x_i$ the element $i$ of
$\mathds{N}$). Thus the modules of the cone are not presented as
\emph{free} modules, but they are isomorphic to modules freely
generated, as it is inferred from the results of Section 3.

The second element of the tuple $(5, 7*x_4 + 8*x_0, 3)$ is represented in Coq
by \texttt{e:= Law (Op (R:= Z\_as\_Ring) 7 (Var \_ (4\%nat: nat\_as\_Setoid)))
       (Op (R:= Z\_as\_Ring) 8 (Var \_ (0\%nat: nat\_as\_Setoid))).}

The required Coq code is then the following:
 {\small\begin{verbatim}
 Eval vm_compute in
  ((Diff(topCC Example_Cone_EffectiveHomology) 2) (5, e, 3)).\end{verbatim}}
\noindent and the result returned by Coq is:
 {\small\begin{verbatim}
 = (-10, Inv e, 5): topCC Example_Cone_EffectiveHomology 2 \end{verbatim}}
\noindent \emph{i.e.},  $(-10, -(7*x_4 + 8*x_0 ), 5)$. If we apply
now the (degree $1$) differential to this element we obtain:
{\small\begin{verbatim}
 = (0, Inv (Inv (Law (Op 7 (Unit Z_as_Ring nat_as_Setoid))
                 (Op 8 (Unit Z_as_Ring nat_as_Setoid)))), 0)
     : topCC Example_Cone_EffectiveHomology 1 \end{verbatim}}
\noindent or, in plain notation, $(0, -(-(7*() + 8*() ), 0)$ which
it is equal (in the setoid) to the null element.
It should be
recalled that our formalization of the free module generated by the
natural numbers directly use the L. Pottier definition for free
modules, and, as a consequence, we are not working with canonical
elements on the free modules or with structures which allow a
reduction to them.

Now, we focus our attention on homotopy operators, that is to say
on morphisms which increase in one unity the degree into the graded
module. We use as ambient structures the chain complexes $Cone(\pi_1)$ and
$Cone(\alpha')$ introduced in the previous section.

Some examples of homotopy operators for $Cone(\alpha')$,
$h=(h_i\colon Cone(\alpha')_i \to Cone(\alpha')_{i+1})_{i \in
\mathds{Z}}$, are the following:
\begin{itemize}
\item $h1= (h1_i)_{i \in \mathds{Z}}$, such that $h1_i(a, b):=(0, a)$,
$(a,b)\in Cone(\alpha')_i$ for all $i \in \mathds{Z}$
\item $h2= (h2_i)_{i \in \mathds{Z}}$, such that $h2_i(a, b):= (b, 0)$,
$(a,b)\in Cone(\alpha')_i$ for all $i \in \mathds{Z}$
\end{itemize}
\noindent Both can be easily implemented in Coq. For
example, the first one is represented through:
 {\small\begin{verbatim}
 Definition h1': forall i:Z, bottomCC Example_Cone_EffectiveHomology i ->
  bottomCC Example_Cone_EffectiveHomology(i + 1):=
   fun (i:Z)(c: bottomCC Example_Cone_EffectiveHomology i) => (0, fst c). \end{verbatim}}

There exist special homotopy operators called \emph{contracting homotopies} which express algorithmically that the chain complex is \emph{acyclic}~\cite{BSM}.
\begin{definition}
A chain complex is \emph{acyclic} if it is possible to define an effective homology from it to the null chain complex.
\end{definition}
\begin{corollary}\label{acyclic-corollary}
Let $CC =(M , d)$ be a chain complex, $CC$ is acyclic if and only if there exists a homotopy operator $h$ defined on $CC$ such that $d \circ h + h \circ d = id$. Such an operator is called \emph{contracting homotopy}.
\end{corollary}

We can test if the previous homotopy operators define a contracting homotopy.  For
instance, the corresponding tactic at degree \texttt{i=1}
choosing as element \mbox{\texttt{(5, 7): bottomCone 2}} for the first
candidate is:
 {\small\begin{verbatim}
 Eval vm_compute in
 (((Diff (bottomCC Example_Cone_EffectiveHomology) 2)[oh](h1 2))[+h]
  ((h1 1)[oh](Diff(bottomCC Example_Cone_EffectiveHomology) 1)))(5, 7). \end{verbatim}}
\noindent resulting in: {\small \texttt{= (0, 0): bottomCC Example\_Cone\_EffectiveHomology 2}}.

For the second homotopy operator over the same element we obtain:
{\small\begin{verbatim}
 Eval vm_compute in
 (((Diff (bottomCC Example_Cone_EffectiveHomology) 2)[oh](h2 2))[+h]
  ((h2 1)[oh](Diff(bottomCC Example_Cone_EffectiveHomology) 1)))(5, 7). \end{verbatim}}
\noindent resulting in: {\small \texttt{= (5, 7): bottomCC Example\_Cone\_EffectiveHomology 2}}.

This means that $h1$ is not a contracting homotopy for
$Cone(\alpha')$. It could be, anyway, acyclic. The homotopy operator
$h2$ could be a candidate for contracting homotopy and, in fact, if
we test other elements in other dimensions we always obtain the
identity.

Moreover, using the homotopy operator $h2$ and the one $h^{Ex}$ in the
effective homology at the end of the previous section, we can define a new
homotopy operator over $Cone(\pi_1)$ with the formula
$h=h^{Ex}+g^{Ex}\circ h2\circ f^{Ex}$. Graphically:
$$
  \xymatrix@C=4pc{Cone(\pi_1)\ar@/^/[rr]^{f^{Ex}}\ar@/4pc/@(r,u)_{h^{Ex}}\ar@(l,d)_{h=h^{Ex}+\ g^{Ex}\circ\  h2\ \circ  f^{Ex}}& &Cone(\alpha') \ar@/^/[ll]^{g^{Ex}}\ar@/4pc/@(l,u)^{h2}
  }
$$
\noindent This homotopy operator can be easily defined in Coq in the following way:
 {\small\begin{verbatim}
 Definition h_topCone:
 (HomotopyOperator(topCC Example_Cone_EffectiveHomology)):=
  fun n: Z => (h_t_t Example_Cone_EffectiveHomology) n [+h]
   (((g_b_t Example_Cone_EffectiveHomology) n) [oh] (h2 n) [oh]
   ((f_t_b Example_Cone_EffectiveHomology) n)). \end{verbatim}}

We can test if it is a candidate to be a contracting homotopy:
 {\small\begin{verbatim}
 Eval vm_compute in
 (((Diff(topCC Example_Cone_EffectiveHomology) 2)[oh](h_topCone 2))
   [+h]((h_topCone 1)[oh]
    ((Diff(topCC Example_Cone_EffectiveHomology) 1))))(5, e, 3). \end{verbatim}}
\noindent whose result is an element equal (in the setoid) to \texttt{(5, e, 3)}.

The testing with other elements and at other degrees is always successful
and this allows us to conjecture that it is really a contracting homotopy.


If that is the case, it could be used to solve a problem that, in
general, is undecidable when working with chain complexes of
infinite type. If an element $x$ is a cycle (that is to say,
$d_n(x)=0$) and the chain complex is acyclic, then there exists an
element $z$ such that $d_{n+1}(z) = x$. Or, in other words, $z$ is a
pre-image of $x$ for the differential. Let us compute such a
pre-image in our example. To this aim, we choose again $x = (-10,
-(7*x_4 + 8*x_0), 5)$ as an element at degree 2. We know already it
is a cycle, because it has been previously computed. Then, if our
homotopy operator $h$ is actually a contracting homotopy, the image
$h(x)$ must be a pre-image of $x$ for $d$ (since $dh(x) + hd(x) =
x$, but $hd(x)=0$). We can test in Coq this fact as follows. First
we apply the homotopy operator on the element:
 {\small\begin{verbatim}
 Eval vm_compute in (h_topCone 2)(-10, Inv e, 5).\end{verbatim}}
\noindent obtaining an element equal to \texttt{(5, e, 0)}. And due to
our previous computations we know that this element is indeed in the
right pre-image because
 {\small\begin{verbatim}
 Eval vm_compute in
 ((Diff(topCC Example_Cone_EffectiveHomology) 2))(5, e, 0). \end{verbatim}}
\noindent gives the required element \texttt{(-10, Inv e, 5)}.

This behaviour is not accidental. The testing is reflecting
a general result relating cones and reductions. Namely:
\begin{proposition}\label{acyclic-proposition} Let $(M, N, f, g, h)$ be a reduction.
Then $Cone(f)$ is an acyclic chain complex.
\end{proposition}
The constructive proof of this proposition gives exactly the
formula we were testing before. Finally, we could proof in Coq that $h2$ and $h$ are indeed contracting homotopies which is now an easy exercise. Also Corollary~\ref{acyclic-corollary} and Proposition~\ref{acyclic-proposition} could be formalized in Coq, although more effort is required. Both tasks are proposed as future work.

\section{Conclusions and Further Work}

In this paper we have presented some examples relating deduction
and computing in the Coq proof assistant. Even if constructive
type theory always allows, in principle, the modeler to execute
terms (by reducing them) this is rarely used in development (or,
at least, it is rarely documented). In our case, testing has been
worked out in an infinite dimensional setting. Concretely, we have
constructed concrete instances of chain complexes of infinite type,
we have computed in Coq with their elements, and we have checked some
formula producing a contracting homotopy on one of the chain complexes.
This testing corresponds to a general theorem that could be, later on,
proved in Coq, too.

The chain complexes of infinite type used as examples in this paper are,
in some sense, artificial. It can be considered as a demonstration of
feasibility. In a future step, we will undertake the implementation
in Coq of more meaningful infinite dimensional spaces. Our first candidates
will be \emph{loop spaces}. The chain complex associated to a combinatorial
loop space (see Kan's $G$ construction in \cite{may}) is of infinite type.
Under good conditions, its homology groups are, however, of finite type.
Computing these homology groups was one of the first challenges solved by
Kenzo (see \cite{BSM}), and working with them in Coq would be an interesting
issue.

One unpleasant aspect of our work is that we are working in a context
where combinations are not in normal form. This implies that, once
a function has been applied, some work is needed to prove
the result is equal to some assumed test value. Several approaches are
known to tackle this reduction to canonical form, and we should systematically
explore some of them to propose a more comfortable way of doing testing in Coq.
Another via to avoid this difficulty could be to give setoids up and
work inside the \emph{ssreflect} framework~\cite{Gonthier}.

Another related line is that of code extraction. We should retake the
works on going from Coq to Scheme~\cite{Letouzey08}, and adapt them to Common
Lisp. Since that we have a model (Kenzo \cite{Kenzo}) of the programs
we would like to extract, the challenge would be to devise Coq statements
and proofs in such a way that the extracted programs would be as close
as possible to the selected Kenzo fragment.

Finally we could study the possibilities of tools like QuickCheck~\cite{QuickCheck} in our setting.
This system allows to test properties of programs automatically by generating a large number of cases (although, up to our knowledge, there is no direct application to Coq code).

\subsubsection*{Acknowledgement} The authors wish to thank the anonymous reviewers for their useful comments.

\end{document}